\documentclass[prl,aps,superscriptaddress,preprint,showpacs]{revtex4}
\usepackage{graphicx}
\usepackage{amssymb}
\begin{document}
\title{Novel $J_{\it{eff}}$ = 1/2 Mott State Induced by Relativistic Spin-Orbit Coupling in Sr$_2$IrO$_4$}

\author{B. J. Kim}
\author{Hosub Jin}
\affiliation{CSCMR \& School of Physics and Astronomy, Seoul
National University, Seoul 151-747, Korea}

\author{S. J. Moon}
\affiliation{ReCOE \& School of Physics and Astronomy, Seoul
National University, Seoul 151-747, Korea}

\author{J.-Y. Kim}
\affiliation{Pohang Accelerator Laboratory, Pohang University of
Science and Technology, Pohang 790-784, Korea}

\author{B.-G. Park}
\affiliation{eSSC \& Department of Physics, Pohang University of
Science and Technology, Pohang 790-784, Korea}

\author{C. S. Leem}
\affiliation{Institute of Physics and Applied Physics, Yonsei
University, Seoul, Korea}

\author{Jaeju Yu}
\affiliation{CSCMR \& School of Physics and Astronomy, Seoul
National University, Seoul 151-747, Korea}

\author{T. W. Noh}
\affiliation{ReCOE \& School of Physics and Astronomy, Seoul
National University, Seoul 151-747, Korea}

\author{C. Kim}
\affiliation{Institute of Physics and Applied Physics, Yonsei
University, Seoul, Korea}

\author{S.-J. Oh}
\affiliation{CSCMR \& School of Physics and Astronomy, Seoul
National University, Seoul 151-747, Korea}

\author{J.-H. Park}
\altaffiliation{Author to whom all correspondences should be
addressed. e-mail: jhp@postech.ac.kr}

\affiliation{Pohang Accelerator Laboratory, Pohang University of
Science and Technology, Pohang 790-784, Korea}

\affiliation{eSSC \& Department of Physics, Pohang University of
Science and Technology, Pohang 790-784, Korea}

\author{V. Durairaj}
\author{G. Cao}
\affiliation{Department of Physics and Astronomy, University of
Kentucky, Lexington, KY 40506}

\author{E. Rotenberg}
\affiliation{Advanced Light Source, Lawrence Berkeley National
Laboratory, Berkeley, CA 96720}

\begin{abstract}

We investigated electronic structure of 5$d$ transition-metal oxide
Sr$_2$IrO$_4$ using angle-resolved photoemission, optical
conductivity, and x-ray absorption measurements and first-principles
band calculations. The system was found to be well described by
novel effective total angular momentum $J_{\it{eff}}$ states, in
which relativistic spin-orbit (SO) coupling is fully taken into
account under a large crystal field. Despite of delocalized Ir 5$d$
states, the $J_{\it{eff}}$-states form so narrow bands that even a
small correlation energy leads to the $J_{\it{eff}}$ = 1/2 Mott
ground state with unique electronic and magnetic behaviors,
suggesting a new class of the $J_{\it{eff}}$ quantum spin driven
correlated-electron phenomena.

\end{abstract}


\maketitle

Mott physics based on the Hubbard Hamiltonian, which is at the root
of various noble physical phenomena such as metal-insulator
transitions, magnetic spin orders, high $T_C$ superconductivity,
colossal magneto-resistance, and quantum criticality, has been
adopted to explain electrical and magnetic properties of various
materials in the last several decades
\cite{Mott,Hubbard,Fazekas,Bednorz,Imada,Fisher}. Great success has
been achieved especially in 3$d$ transition-metal oxides (TMOs), in
which the 3$d$ states are well localized to yield strongly
correlated narrow bands with a large on-site Coulomb repulsion $U$
and a small band width $W$. As predicted, most stoichiometric 3$d$
TMOs are antiferromagnetic Mott insulators \cite{Imada}. On the
other hand, 4$d$ and 5$d$ TMOs were considered as weakly-correlated
wide band systems since $U$ is largely reduced due to delocalized
4$d$ and 5$d$ states \cite{Ryden}. Anomalous insulating behaviors
were recently reported in some 4$d$ and 5$d$ TMOs
\cite{Nakatsuji,Lee,Crawford,Mandrus,Cava}, and the importance of
correlation effects was recognized in 4$d$ TMOs such as
Ca$_2$RuO$_4$ and Y$_2$Ru$_2$O$_7$, which were interpreted as Mott
insulators near the border line of the Mott criteria, \textit{i.e.}
$U \sim W$ \cite{Nakatsuji,Lee}. However, as 5$d$ states are
spatially more extended and $U$ is expected to be further reduced,
insulating behaviors in 5$d$ TMOs such as Sr$_2$IrO$_4$ and
Cd$_2$Os$_2$O$_7$ have been puzzling \cite{Crawford,Mandrus}.

Sr$_2$IrO$_4$ crystallizes in the K$_2$NiF$_4$ layered structure as
La$_2$CuO$_4$ and its 4$d$ counterpart Sr$_2$RhO$_4$
\cite{Crawford,Vogt}. Considering its odd number of electrons per
unit formula (5$d^5$), one expects a metallic state in a na\"{\i}ve
band picture. Indeed Sr$_2$RhO$_4$ (4$d^5$) is a normal metal. Its
Fermi surface (FS) measured by the angle resolved photoemission
spectroscopy (ARPES) agrees well with the first-principles band
calculation results \cite{Kim,Baumberger}. Since both $d^5$ systems
have identical atomic arrangements with nearly the same lattice
constants and bond angles \cite{Crawford,Vogt}, one expects almost
the same FS topology. Sr$_2$IrO$_4$, however, is unexpectedly an
insulator with weak ferromagnetism \cite{Crawford}. At this point,
it is natural to consider the spin-orbit (SO) coupling as a
candidate responsible for the insulating nature since its energy is
much larger than that in 3$d$ and 4$d$ systems. Recent band
calculations showed that the electronic states near $E_F$ can be
modified considerably by the SO coupling in 5$d$ systems, and
suggest a new possibility of the Mott instability \cite{Singh}. It
indicates that the correlation effects can be important even in 5$d$
TMOs when combined with strong SO coupling.

In this Letter, we show formation of new quantum state bands with
effective total angular momentum $J_{\it{eff}}$ in 5$d$ electron
systems under a large crystal field, in which the SO coupling is
fully taken into account, and also report for the first time
manifestation of a novel $J_{\it{eff}} = 1/2$ Mott ground state
realized in Sr$_2$IrO$_4$ by using ARPES, optical conductivity, and
x-ray absorption spectroscopy (XAS) and first-principles band
calculations. This new Mott ground state exhibits novel electronic
and magnetic behaviors; for examples, spin-orbit integrated narrow
bands and an exotic orbital dominated local magnetic moment,
suggesting a new class of the $J_{\it{eff}}$ quantum spin driven
correlated-electron phenomena.

Single crystals of Sr$_2$IrO$_4$ were grown by flux method
\cite{Cao}. ARPES spectra were obtained from cleaved surfaces
\textit{in situ} under vacuum of 1x10$^{-11}$ Torr at the beamline
7.0.1 of the Advanced Light Source with $h\nu$ = 85 eV and $\Delta
E$ = 30 meV. The temperature was maintained at 100 K to avoid
charging. The chemical potential $\mu$ was referred to $E_F$ of a
clean Au electrically connected to the samples. The electronic
structure calculations were performed by using first-principles
density-functional-theory codes with LDA and LDA + $U$ methods
\cite{Han}. The optical reflectivity $R(\omega)$ was measured at 100
K between 5 meV and 30 eV and the optical conductivity
$\sigma(\omega)$ was obtained by using Kramers-Kronig (KK)
transformation. The validity of KK analysis was checked by
independent ellipsometry measurements between 0.6 and 6.4 eV. The
XAS spectra were obtained at 80 K under vacuum of 5x10$^{-10}$ Torr
at the Beamline 2A of the Pohang Light Source with $\Delta h \nu$ =
0.1 eV.

Here we propose a schematic model for emergence of a novel Mott
ground state by a large SO coupling energy $\zeta_{SO}$ as shown in
Fig. 1. Under the $O_h$ site symmetry the 5$d$ states are split into
triplet $t_{2g}$ and doublet $e_g$ orbital states by the crystal
field energy 10Dq. In general, 4$d$ and 5$d$ TMOs have sufficiently
large 10Dq to yield $t_{2g}^5$ low-spin ionic state for
Sr$_2$IrO$_4$, and thus the system with partially filled wide
$t_{2g}$ band would become a metal (Fig. 1(a)). An unrealistically
large $U \gg W$ could lead to a typical spin $S = 1/2$ Mott
insulator (Fig. 1(b)). However, as $U$ is expected to be even
smaller than that in 4$d$ TMOs, a reasonable $U$ cannot lead to an
insulating state as seen from the fact that Sr$_2$RhO$_4$ is a
normal metal. Now let us take the SO coupling into account. The
$t_{2g}$ states effectively correspond to the orbital angular
momentum $L$ = 1 states with $\psi_{m_l=\pm1}=\mp(|zx\rangle \pm
i|yz\rangle)/\sqrt{2}$ and $\psi_{m_l=0}=|xy\rangle$. In the
\textit{strong SO coupling limit}, the $t_{2g}$ band splits into
\textit{effective} total angular momentum $J_{\it{eff}} = 1/2$
doublet and $J_{\it{eff}} = 3/2$ quartet bands (Fig. 1(c)). Note
that the $J_{\it{eff}} = 1/2$ is energetically higher than the
$J_{\it{eff}} = 3/2$, seemingly against the Hund's rule, since the
$J_{\it{eff}} = 1/2$ branches out from the $J_{5/2}$ (5$d_{5/2}$)
manifold due to the large crystal field as depicted in Fig. 1(e). As
a result, with four electrons filling up the $J_{\it{eff}} = 3/2$
band and the remaining one in the $J_{\it{eff}} = 1/2$ band, the
system is effectively reduced to a half-filled $J_{\it{eff}} = 1/2$
single band system (Fig. 1(c)). The $J_{\it{eff}} = 1/2$ spin-orbit
integrated states form a narrow band so that relatively small $U$
opens a Mott gap, making it a $J_{\it{eff}} = 1/2$ Mott insulator
(Fig. 1(d)). The narrow band width is due to reduced hopping
elements of the $J_{\it{eff}} = 1/2$ states with isotropic orbital
and mixed spin characters, which are differentiated from the atomic
$J = 1/2$ states as discussed later in detail. The formation of the
$J_{\it{eff}}$ bands due to the large $\zeta_{SO}$ explains why
Sr$_2$IrO$_4$ ($\zeta_{SO}\sim$ 0.4 eV, estimated from the
calculation) is insulating while Sr$_2$RhO$_4$ ($\zeta_{SO}\sim$
0.15 eV) is metallic.

The $J_{\it{eff}}$ band formation is well-justified in the first
principles band calculations on Sr$_2$IrO$_4$ based on local-density
approximation (LDA) as well as LDA + $U$ with and without including
the SO coupling presented in Fig. 2. The LDA result (Fig. 2(a))
yields a metal with a wide $t_{2g}$ band as shown in Fig. 1a, and
the Fermi surface (FS) is nearly identical to that of Sr$_2$RhO$_4$
\cite{Kim,Baumberger}. The FS, composed of one-dimensional $yz$ and
$zx$ bands, is represented by hole-like $\alpha$ and $\beta_X$
sheets and an electron-like $\beta_M$ sheet centred at $\Gamma$, X,
and M points, respectively \cite{Kim}. As the SO coupling is
included (Fig. 2(b)), the FS becomes rounded but retains the overall
topology. Despite small variations in the FS topology, the the band
structure changes remarkablly: Two narrow bands are split off near
$E_F$ from the rest of the bands due to formation of the
$J_{\it{eff}}$ = 1/2 and 3/2 bands as shown in Fig. 1(c). The
circular shaped FS reflects the isotropic orbital character of the
$J_{\it{eff}} = 1/2$ states.

Such a narrow band near $E_F$ suggests that a small $U$ can drive
the system to a Mott instability. Indeed, a modest $U$ value opens
up a Mott gap and splits the half-filled $J_{\it{eff}} = 1/2$ band
into the upper (\textit{UHB}) and lower Hubbard bands
(\textit{LHB}), as presented in Fig. 1(d). The \textit{full} LDA +
SO + $U$ results shown in Fig. 2(c) manifest the $J_{\it{eff}} =
1/2$ Mott state. In comparing the LDA + SO and LDA + SO + $U$
results (Fig. 2(b),(c)), one can see that the band gap is opened up
by simply shifting up the electron-like M sheet and down the
hole-like $\Gamma$ and X sheets, yielding a valence band maxima
topology as shown in the left panel of Fig. 2(c). It must be
emphasized that \textit{without the SO coupling, LDA + $U$ alone can
not account for the band gap}. In that case, the FS topology changes
only slightly from the LDA one, as shown in Fig. 2(d), because $W$
is so large that the small $U$ can not play a major role. This
result demonstrates that the strong SO coupling is essential to
trigger the Mott transition in Sr$_2$IrO$_4$, which reduces to a
$J_{\it{eff}} = 1/2$ Hubbard system.

The electronic structure predicted by the LDA + SO + $U$ is borne
out by ARPES results presented in Fig. 3. The energy distribution
curves (EDCs) near $\mu$ display dispersive band features, none of
which crosses over $\mu$ as expected in an insulator. Fig. 3(b)-(d)
show intensity maps at binding energies of $E_B$ = 0.2, 0.3, and 0.4
eV, highlighting the evolution of the electronic structure near
$\mu$. The first valence band maximum ($\beta_X$) appears at the X
points (Fig. 3(b)). As $E_B$ increases (Fig. 3(c),(d)), another band
maximum ($\alpha$) appears at the $\Gamma$ points. The band maxima
can also be ascertained in EDCs (Fig. 3(a)). The observed electronic
structure agrees well with the LDA + SO + $U$ results, reproducing
the correct valence band maxima topology (see the left panel of Fig.
2(c)). Remarkably, the topmost valence band, which represents the
$J_{\it{eff}} = 1/2$ \textit{LHB}, has a very small dispersion
($\sim$ 0.5 eV) although the 5$d$ states are spatially extended and
strongly hybridized with the O 2$p$ states.

The band calculations and ARPES provide solid evidences for the
$J_{\it{eff}}$ Mott picture. The unusual electronic character of the
$J_{\it{eff}} = 1/2$ Mott state is further confirmed in the optical
conductivity \cite{Moon} and the O 1$s$ polarization dependent XAS.
The optical conductivity in Fig. 4(a), which shows an $\sim$ 0.1 eV
insulating gap in good agreement with the observed resistivity with
an activation energy of 70 meV \cite{Kim2}, displays an uncommon
double-peak feature with a sharp peak A around 0.5 eV and a rather
broad peak B around 1 eV. Considering the delocalized 5$d$ states,
it is rather unusual to have such a sharp peak A whose width is much
smaller than that of the peaks in 3$d$ TMOs. However, this feature
is a natural consequence of the $J_{\it{eff}}$-manifold Hubbard
model as depicted in Fig. 1(d). The transitions within the
$J_{\it{eff}} = 1/2$ manifold, from \textit{LHB} to \textit{UHB},
and from $J_{\it{eff}} = 3/2$ to the $J_{\it{eff}} = 1/2$
\textit{UHB} results in the sharp peak A and a rather broad peak B,
respectively. A direct evidence of the $J_{\it{eff}} = 1/2$ state
comes from the XAS which enables one to characterize the orbital
components by virtue of the strict selection rules\cite{Mizokawa}.
The results in Fig. 4(b) show an orbital ratio $xy$ : $yz$ : $zx$ =
1 : 1 : 1 within an estimation error ($<$ 10 \%) for the unoccupied
$t_{2g}$ state. In the ionic limit, the $J_{\it{eff}} = 1/2$ states
are $|J_{\it{eff}} = 1/2, m_{J_{\it{eff}}}=\pm
1/2\rangle=(|zx,\pm\sigma\rangle\pm
i|yz,\pm\sigma\rangle\mp|xy,\mp\sigma\rangle)/\sqrt{3}$, where
$\sigma$ denotes the spin state. In the lattice, the inter-site
hopping, the tetragonal and rotational lattice distortions, and
residual interactions with $e_g$ manifold could contribute to
off-diagonal mixing between the ionic $J_{\it{eff}}$ states.
However, the mixing seems to be minimal and the observed isotropic
orbital ratio, which is also predicted in the LDA + SO + $U$,
validates the $J_{\it{eff}} = 1/2$ state.

The $J_{\it{eff}} = 1/2$ Mott state contributes to unusual magnetic
behaviors. The total magnetic moment is dominated by the orbital
moment. In the ionic limit, the spin state of the $J_{\it{eff}} =
1/2$ state is a mixture of $\sigma$ (up spin) and $-\sigma$ (down
spin) and yields $|\langle S_z\rangle|= 1/6$. Meanwhile the orbital
state yields $|\langle L_z\rangle|= 2/3$, resulting in twice larger
orbital moment than the spin one, \textit{i.e.} $|\langle L_z
\rangle|= 2|\langle 2S_z \rangle|$. Note that the $J_{\it{eff}} =
1/2$ is distinguished from the atomic $J = 1/2$ ($|L-S|$) with $L$
=1 and $S$ = 1/2 despite the formal equivalence. The $J$ = 1/2 has
total magnetic moment $\langle L_z + 2S_z\rangle = \pm 1/3$ with
opposite spin and orbital direction ($L-S$), while the $J_{\it{eff}}
= 1/2$ gives $\langle L_z + 2S_z\rangle = \pm 1$ with parallel one.
The $J_{\it{eff}} = 1/2$ ($|L_{\it{eff}} - S|$) is exactly analogous
to the $J$ = 1/2 ($|L - S|$) with mapping $L_{\it{eff,z}}\rightarrow
- L_z$. This is because the $J_{\it{eff}} = 1/2$ is branched off
from the atomic $J$ = 5/2 manifold ($L+S$) by the crystal field, the
same reason for the violation of the Hund's rule (Fig. 1(e)). This
aspect differentiates 5$d$ TMOs from 3$d$ TMOs described by
spin-only moments and also from rare-earth compounds with
atomic-like $J$ states.

The LDA + SO + \textit{U} predicts that the magnetic ground state
has weak ferromagnetism resulting from canted antiferromagnetic
(AFM) order with approximately 22$^\circ$ canting angle in the
plane. The predicted local moment at each Ir site is 0.36 $\mu_B$
with 0.10 $\mu_B$ spin and 0.26 $\mu_B$ orbital contributions. This
value is only about one-third of the ionic value 1 $\mu_B$ (0.33
$\mu_B$ from the spin and 0.67 $\mu_B$ from the orbital) for
$J_{\it{eff}} = 1/2$ but still retains the respective ratio close to
1:2. This large reduction, however, seems to be natural as it is
considered that the Ir 5$d$ orbitals strongly hybridize with
neighboring O 2$p$ ones and thus significant parts of the moments
are canceled in the AFM order. Indeed, Sr$_2$IrO$_4$ shows weak
ferromagnetism below the Curie temperature with local moment
$\mu_{\it{eff}} = 0.5 \mu_B$/Ir, about one-third of $\mu_{\it{eff}}
= 1.73 \mu_B$ for $S$ = 1/2, as determined from the magnetic
susceptibility above $T_C$ \cite{Cao}. It should be noted that the
origin of the canted AFM order is different from that in the
spin-based Mott insulators, which is attributed to the spin canting
due to the Dzyaloshinskii-Moriya (DM) interaction
\cite{Dzyloshinskii}, the first order perturbation term of the SO
coupling on the S basis states. But the $J_{\it{eff}}$ states, in
which the SO coupling is fully included, are free from the DM
interaction. Their canted AFM order should be explained by the
lattice distortion, and the canting angle is indeed nearly identical
to that in the Ir-O-Ir bond \cite{Crawford}.

The peculiar electronic and magnetic properties of Sr$_2$IrO$_4$ can
be understood as characteristics of the $J_{\it{eff}} = 1/2$ Mott
insulator. In spite of the extended 5$d$ states with a small $U$,
narrow Hubbard bands with a new quantum number $J_{\it{eff}}$
different from the ordinary atomic $J$ emerge through the strong SO
coupling under the large crystal field. This suggests a new class of
materials, namely {\it{spin-orbit integrated narrow band system}}.
The $J_{\it{eff}} = 1/2$ quantum "spin", which incorporates the
orbital degrees of freedom, is expected to bring in new quantum
behaviors. Indeed, recent findings show that many iridates display
highly unusual behaviors, for examples, non-Fermi liquid behaviors
in SrIrO$_3$ \cite{Cao2} and a spin liquid ground state in
Na$_4$Ir$_3$O$_8$ \cite{Okamoto}. With the relativistic SO coupling,
the system is in a new balance of the spin, orbital, and lattice
degrees of freedom. It implies that the underlying physics of 5$d$
TMOs is not a simple adiabatic continuation of the 3$d$ TMO physics
to a small $U$ regime and a new paradigm is required for
understanding their own novel phenomena. "What novel phenomena
emerge in the vicinity of this new Mott insulator?" remains as an
open question.

Authors thank T. Arima and H. Takagi for invaluable discussions.
This work was supported by MOST/KOSEF through eSSC at POSTECH, CSCMR
at SNU, ReCOE Creative Research Initiative Program, under grant
R01-2007-000-11188-0, POSTECH research fund, and BK21 program. The
work at UK was supported by an NSF grant DMR-0552267. The
experiments at ALS and PLS are supported by the DOE Office of Basic
Energy Science and in part by MOST and POSTECH, respectively.


\newpage
\begin{figure}
\begin{center}
\includegraphics[width=0.85\textwidth]
{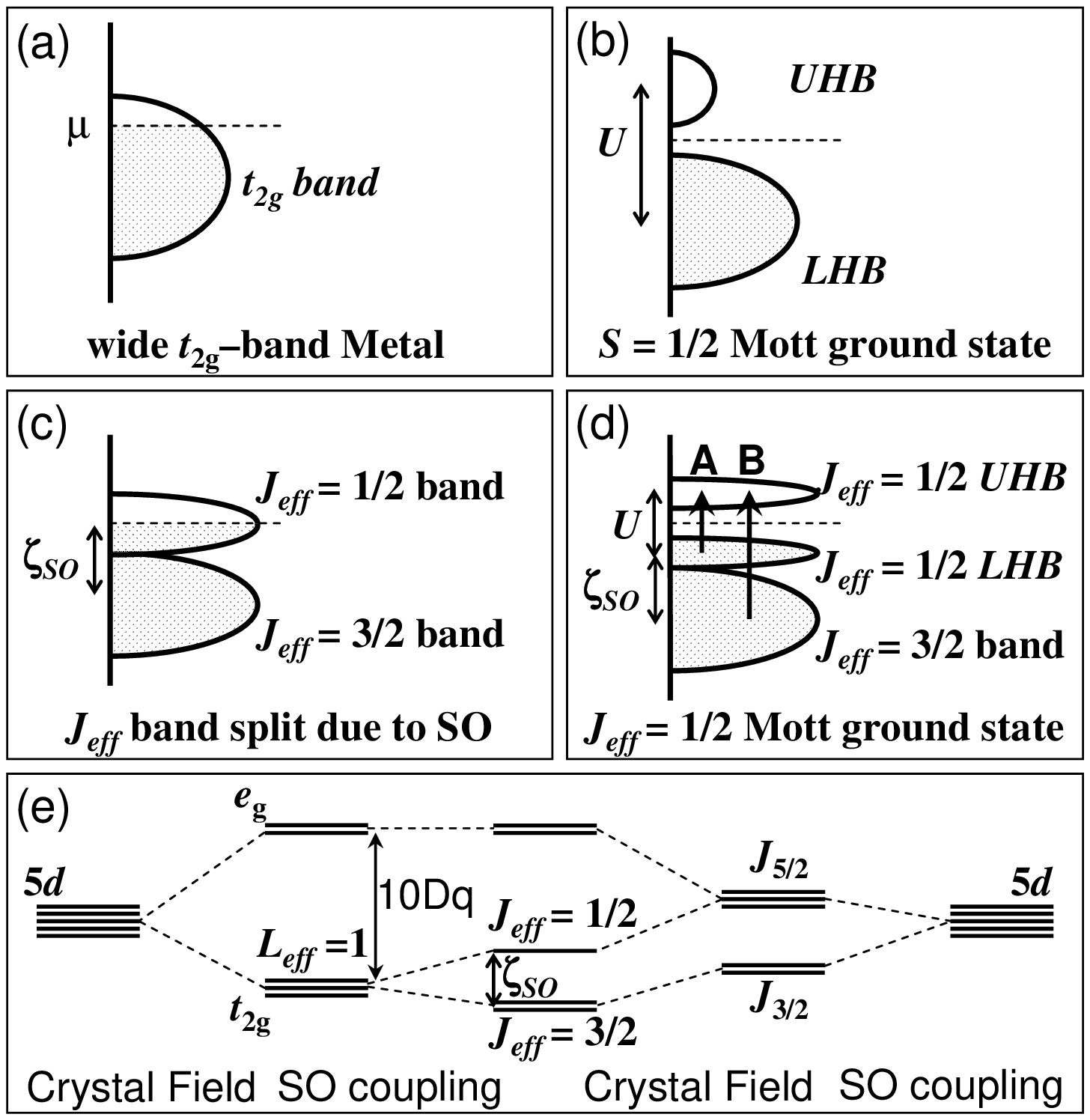} \caption{Schematic energy diagrams for the 5$d^5$
($t_{2g}^5$) configuration (a) without SO and Hubbard $U$, (b) with
an unrealistically large $U$ but no SO, (c) with SO but no $U$, and
(d) with both SO and $U$. Possible optical transitions A and B are
indicated by arrows. (e) 5$d$ level splittings by the crystal field
and SO coupling.} \label{Fig1}
\end{center}
\end{figure}
\newpage
\begin{figure}
\begin{center}
\includegraphics[width=0.85\textwidth]
{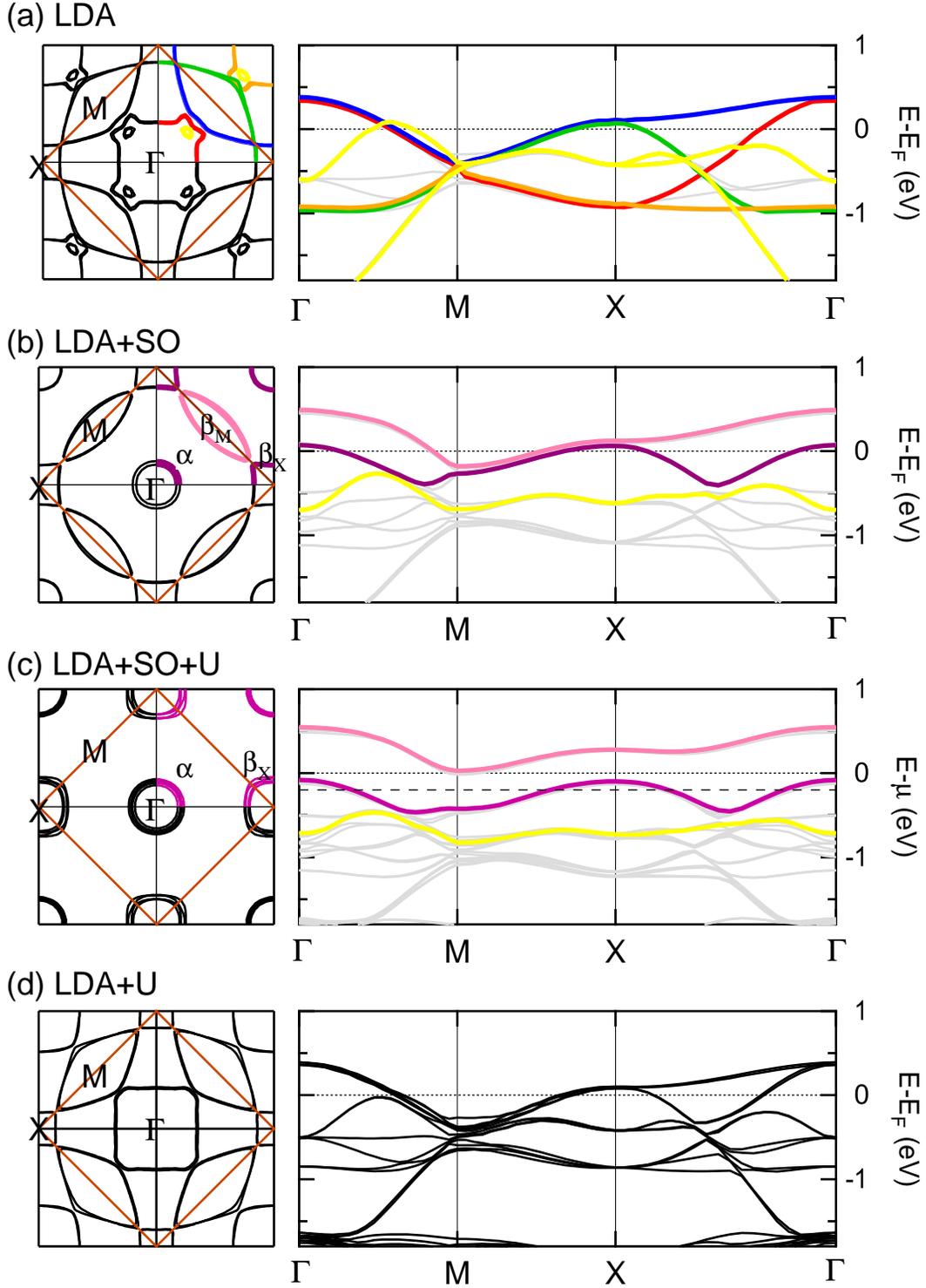} \caption{(Color Online) Theoretical Fermi surfaces and
band dispersions in (a) LDA, (b) LDA + SO, (c) LDA + SO + $U$ (2
eV), and (d) LDA + $U$. In (c), the left panel shows topology of
valence band maxima ($E_B$ = 0.2 eV) instead of the FS. See the
text.} \label{Fig2}
\end{center}
\end{figure}
\newpage
\begin{figure}
\begin{center}
\includegraphics[width=0.85\textwidth]
{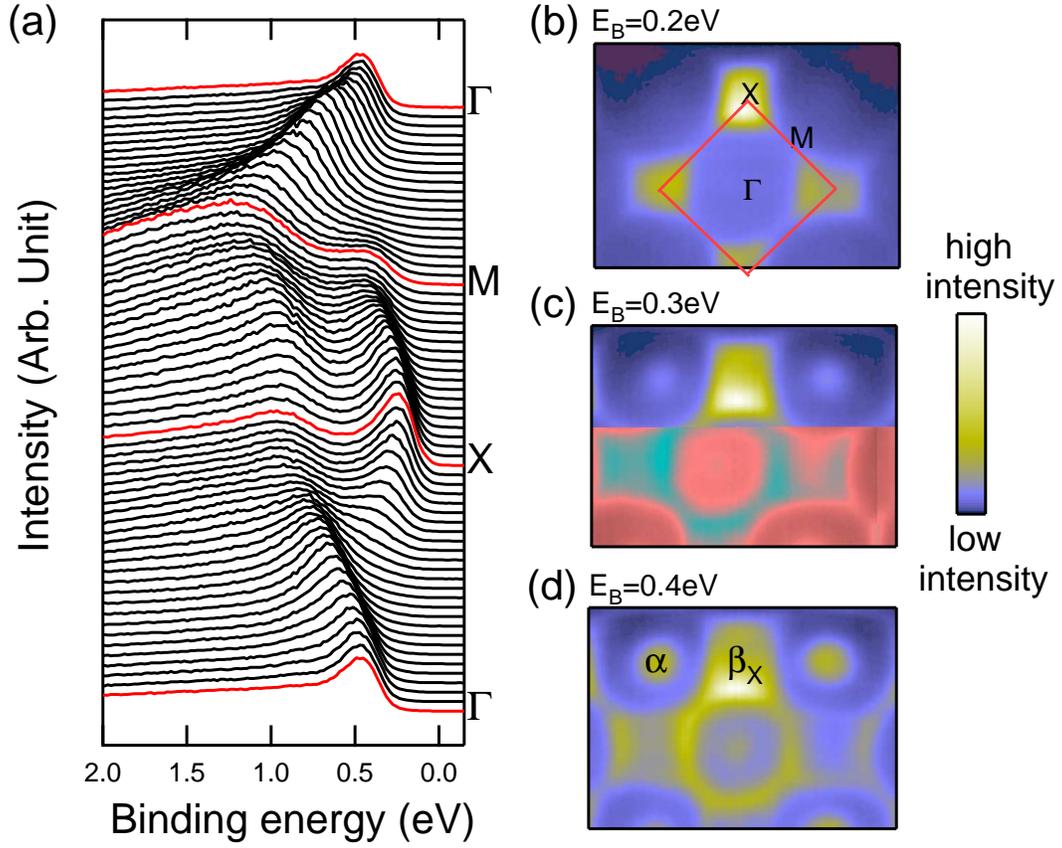} \caption{(Color Online) (a) EDCs up to binding energy
$E_B$ = 2 eV along high symmetry lines. (b)-(d) ARPES intensity maps
at $E_B$ = 0.2 eV, 0.3 eV, and 0.4 eV. Brillouin zone (small square)
is reduced from the original one due to the $\sqrt{2}\times\sqrt{2}$
distortion.} \label{Fig3}
\end{center}
\end{figure}
\newpage
\begin{figure}
\begin{center}
\includegraphics[width=0.85\textwidth]
{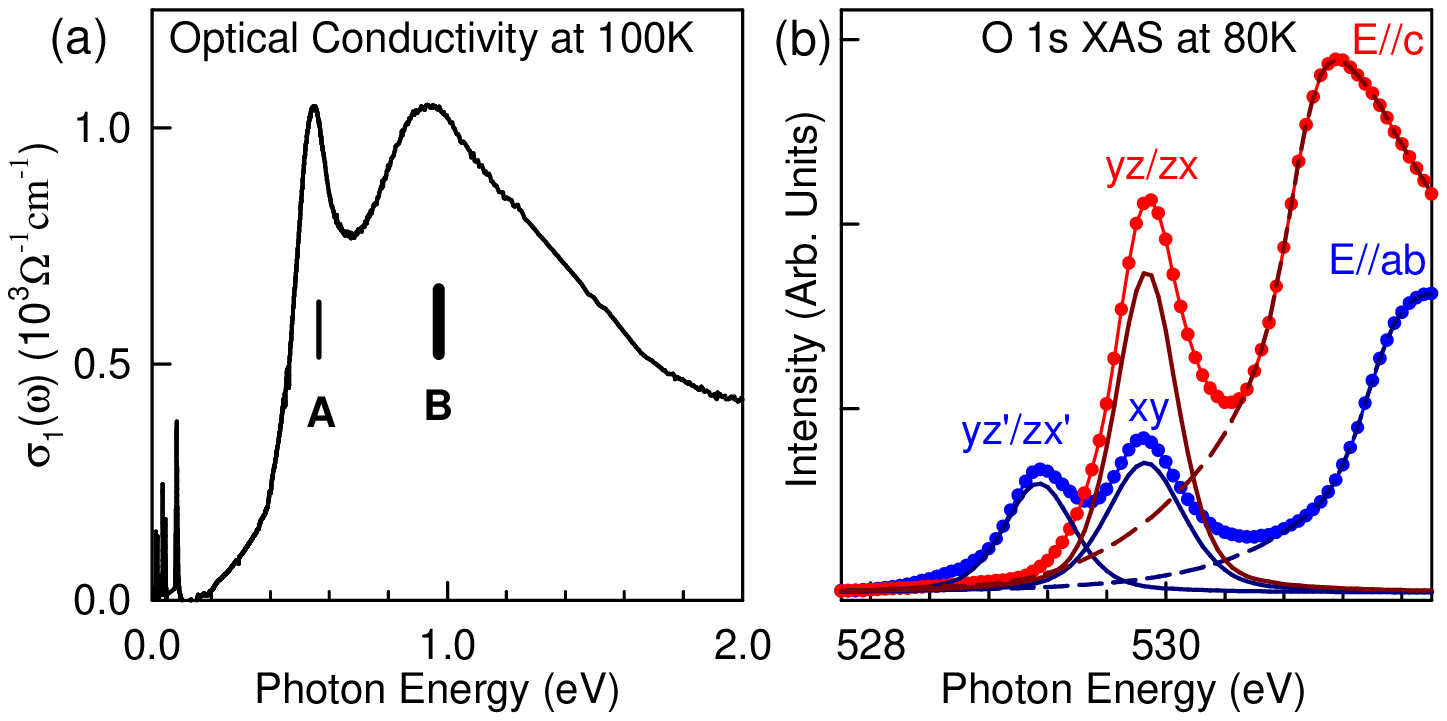} \caption{(Color Online) (a) Optical conductivity. Peak A
and B corresponds to transitions denoted in Fig. 1(d). (b) The O
1$s$ polarization dependent XAS spectra (dotted lines) compared with
expected spectra (solid lines) under an assumption of $xy$:$yz$:$zx$
= 1:1:1 ratio. $xy$ and $yz/zx$ denote transitions from in-plane
oxygens while $yz^\prime/zx^\prime$ from apical oxygens, and the
energy difference corresponds to their different O 1$s$ core-hole
energies \cite{Mizokawa}.} \label{Fig4}
\end{center}
\end{figure}

\end{document}